\documentclass[fleqn,usenatbib]{mnras}
\usepackage[T1]{fontenc}
\usepackage{ae,aecompl}

\usepackage{graphicx}	
\usepackage{amsmath}	
\usepackage{amssymb}	
\usepackage{float}

\title[RadioAstron observations of PSR B0329+54]{PSR B0329+54: Substructure in the scatter-broadened image discovered with RadioAstron on baselines up to 330,000 km}

\author[M. V. Popov et al.]
{Mikhail V.\ Popov$^{1}$,
Norbert Bartel$^{2}$,
Carl R. Gwinn$^{3}$,
Michael D. Johnson$^{4}$,
\newauthor
Andrey Andrianov$^{1}$, 
Evgeny Fadeev$^{1}$, 
Bhal Chandra Joshi$^{5}$, 
Nikolay Kardashev$^{1}$,
\newauthor
Ramesh Karuppusamy$^{6}$,
Yuri Y. Kovalev$^{1,6}$\thanks{E-mail: yyk@asc.rssi.ru},
Michael Kramer$^{6}$,
Alexey Rudnitskiy$^{1}$,
\newauthor
Vladimir Shishov$^{7}$,
Tatiana Smirnova$^{7}$,
Vladimir A. Soglasnov$^{1}$, 
J. Anton Zensus$^{6}$
\\
$^{1}$P.~N.~Lebedev Physical Institute, Astro Space Center, Profsoyuznaya 84/32, Moscow 117997, Russia\\
$^{2}$York University, 4700 Keele St, Toronto, ON M3J 1P3, Canada\\
$^{3}$University of California, Santa Barbara, CA 93106, USA\\
$^{4}$Harvard-Smithsonian Center for Astrophysics, 60 Garden St, Cambridge, MA 02138, USA\\
$^{5}$National Centre for Radio Astrophysics, Post Bag 3, Ganeshkhind, Pune 411007, India\\
$^{6}$Max-Planck-Institut f\"ur Radioastronomie, Auf dem H\"ugel 69, Bonn 53121, Germany\\
$^{7}$P.~N.~Lebedev Physical Institute, Pushchino Radio Astronomy Observatory, Pushchino 142290, Moscow region, Russia
}

\date{Accepted 2016 September 13. Received 2016 September 13; in original form 2016 July 23}

\pubyear{2016}

\begin{document}
\label{firstpage}
\pagerange{\pageref{firstpage}--\pageref{lastpage}}
\maketitle

\begin{abstract}
We have resolved the scatter-broadened image of PSR~B0329+54 and detected substructure within it. These results are not influenced by any extended structure of a source but instead are directly attributed to the interstellar medium. We obtained these results at 324~MHz with the ground-space interferometer RadioAstron which included the space radio telescope (SRT), ground-based Westerbork Synthesis Radio Telescope and 64-m Kalyazin Radio Telescope on baseline projections up to 330,000~km in 2013~November~22 and 2014~January~1~to~2. At short 15,000 to 35,000~km ground-space baseline projections the visibility amplitude decreases with baseline length providing a direct measurement of the size of the scattering disk of $4.8\pm0.8$~mas. At longer baselines no visibility detections from the scattering disk would be expected. However, significant detections were obtained with visibility amplitudes of~3~to~5\% of the maximum scattered around a mean and approximately constant up to 330,000~km. These visibilities reflect substructure from scattering in the interstellar medium and offer a new probe of ionized interstellar material. The size of the diffraction spot near Earth is $17,000\pm3,000$~km. With the assumption of turbulent irregularities in the plasma of the interstellar medium, we estimate that the effective scattering screen is located $0.6\pm0.1$ of the distance from Earth toward the pulsar.
\end{abstract}


\section{Introduction}

All images of radio sources from outside our solar system are influenced by scattering in the interstellar medium (ISM) of our Galaxy. Determining the properties of the scattering is essential for studying the characteristics of the ISM and for a proper interpretation of astronomical radio observations. Pulsars are an almost ideal type of a celestial source for such studies. They are almost point-like so that the results of the study are not influenced by the structure of the celestial source but are almost completely attributable to the influence of the plasma turbulence of the ISM. Scattering of the pulsar signal in the ISM results in angular broadening of the pulsar image, temporal broadening of the pulses, modulation or scintillation of their intensities and, through diffraction patterns, distortion of their radio spectra. Scattering effects of the ISM have been the subject of studies by several authors. Theoretical studies were made by, e.g., \citet{prokhorov1975, rickett1977,goodman1989, narayan1989, gwinn1998}. Observational studies were made with the ground-based VLBI for extragalactic radio sources \citep[e.g.,][]{ojha2006,lazio2008,pus2013,PK2016}, the galactic center Sgr~A$^*$~\citep[e.g.,][]{Lo1998,Bower2006,gwinn2014,PK2016}, and pulsars \citep[e.g.,][]{kondratiev2007,gwinn1993,desai1992,bartel1985}.

PSR~B0329+54 is the brightest pulsar below 1~GHz in the northern hemisphere. With a galactic longitude of $145^\circ$ and latitude of $-1\fdg2$, and at a parallax distance of $1.03^{+0.13}_{-0.12}$~kpc \citep{brisken2002}, the pulsar is located just at the outer edge of the Orion spiral arm. The scattering disk remained unresolved. An early upper limit of the angular size at 2.3~GHz is $\theta_\mathrm{scat}<1$~mas \citep{bartel1985}.

Extending baselines into space, the first ground-space VLBI observations of a pulsar were made with the VLBI Space Observatory Program (VSOP) by \cite{yangalov2001}. The observed pulsar was PSR~B0329+54. With the VSOP observations its scattering disk was not resolved. The upper limit of the angular size at 1.7~GHz was determined to be $\theta_\mathrm{scat}<2$~mas \citep{yangalov2001}. However, the VSOP observations were also done at an insufficiently low frequency  and with baselines only up to about 25,000~km. Ground-space VLBI with RadioAstron allows observations at a frequency as low as 324~MHz, where scattering effects are expected to be much stronger and with baselines about 15 times longer. Since the scattering size increases as $\nu^{-2}$, the size of $\theta_\mathrm{scat}<2$~mas at 1.7~GHz and the size of $\theta_\mathrm{scat} < 1$~mas at 2.3~GHz from the ground-ground VLBI observations correspond to a size of $<50$~mas at 324~MHz.  At this frequency RadioAstron has an angular resolution of about 1~mas at its longest baselines. Therefore observations with RadioAstron have the potential to resolve PSR~B0329+54's scattering disk and perhaps reveal hitherto unknown structure in the ISM. Here we report on such investigations with RadioAstron with baselines up to 330,000~km at a frequency of 324~MHz.

This paper is the second one in a series. The first paper \citep[Paper I,][]{gwinn+2016} gives a mathematical description of the functions obtained from the interferometer observations and leading to an evaluation of the scattering structure in the interstellar medium along the line of sight to PSR~B0329+54. In this second paper we focus on the visibility magnitude as a function of projected baseline length and on a comparison between angular and temporal broadening. This will allow us to draw conclusions about
the size of the scatter-broadened image of PSR~B0329+54 as well as the characteristics of the diffraction spot near Earth and the distance of the scattering screen.

\section{Observations}\label{obs} 

The observations being analysed in this paper were made in two sessions, on 2013 November 22, and 2014 January 1-2 with the 14$\times$25-m Westerbork Synthesis Radio Telescope (WB), the 64-m Kalyazin Radio Telescope (KL), and the 10-m Space Radio Telescope (SRT) on projected baselines from 17,000 km to 330,000 km. The RadioAstron ground-space interferometer operated during six periods of 80 to 120 min with gaps in between necessitated by thermal constraints on the ``Spektr-R'' spacecraft.
Additional RadioAstron observations were made in an earlier session, on 2012 November 26-29, with the 110-m Robert C.\ Byrd Green Bank Telescope (GBT) on projected baselines from 56,000 to 235,000~km.
The 2012 and 2014 observations were discussed in Paper~I. 
The main scattering parameters derived from the 2012 experiment, such as the decorrelation bandwidth and the scattering time (see section~\ref{IntVis} for definitions), differ by about two times from those for the other observing sessions, and may not be used in a common analysis.
We therefore apply the data from the 2012 SRT-GBT experiment in this paper only for purposes of checking SRT calibration as well as determining the average profile of a pulse and its phase in the pulsar period.
We summarize the 2013 and 2014 observations in Table~\ref{tab:obs} of this paper.

\begin{table*}
\centering

	\caption{RadioAstron Space Radio Telescope, Kalyazin, WSRT observations log}
	\label{tab:obs}
	\begin{tabular}{c c c c c c c c} 
		\hline
		\hline
Epoch & Start of obs. & Length of obs.$^{a}$ & Length of scans$^{b}$ & Number of scans & {b}$^{c}$ & Telescopes$^{d}$ & Polarization$^{e}$ \\
              & (UT, hh:mm)   & (min) & (s) & &(M$\lambda$) &  &\\
		\hline
		2013 Nov 22  & 01:10 & 120 & 1170 &  6 &  343 & KL, SRT & RCP + LCP\\
		2013 Nov 22  & 07:00 & 120 & 1170 &  4 &  350 & KL, SRT & RCP + LCP\\
		2013 Nov 22  & 11:40 & 100 & 1170 &  5 &  356 & KL, SRT & RCP + LCP\\
		2014 Jan 01  & 21:00 & 320 & 1170 & 17 &    2 & WB, KL & RCP\\
		2014 Jan 01  & 23:10 & 80  & 1170 &  4 &   24 & WB, SRT & RCP\\
		2014 Jan 01  & 23:10 & 80  & 1170 &  4 &   26 & KL, SRT & RCP\\
		2014 Jan 02  & 02:40 & 120 & 1170 &  6 &   71 & KL, SRT & RCP\\
		2014 Jan 02  & 07:00 & 120 & 1170 &  5 &  111 & KL, SRT & RCP\\
		\hline
	\end{tabular}

\raggedright
Note: In addition to the shown epochs, SRT-GBT data observed in 2012 \citep{gwinn+2016} were used for purposes of checking SRT calibration as well as determining the average profile of a pulse and its phase in the pulsar period.\\
$^{a}$ Length of observations from the start time;\\
$^{b}$ After each scan of 1170~s, there is a 30~s gap;\\
$^{c}$ Mean projected baseline length;\\
$^{d}$ WB (14 $\times$ 25-m Westerbork Synthesis Radio Telescope), KL (64-m Kalyazin Radio Telescope), SRT (10-m RadioAstron Space Radio Telescope);\\
$^{e}$ Recorded polarization, RCP (right circular polarization), LCP (left circular polarization).
\end{table*}

In all observations only the upper
sideband of the 316-332~MHz band was recorded, with one bit digitization at the
RadioAstron space radio telescope (SRT) and with two bit digitization at the WB and KL. The auto-level gain control (AGC), phase-cal, and noise
diode were turned off during our observations to avoid interference with pulses from
the pulsar. The data from the observations with SRT were transmitted in real time to the
telemetry station in Pushchino \citep{kardashev2013}. For the recording of the data we used the RadioAstron data recorder (RDR) at Pushchino and KL and the MkVb recording system at WSRT. The recorded data were then transferred via internet to the
Astro Space Center (ASC) in Moscow for processing by the ASC correlator.

\section{Data reduction}
\label{red}

\subsection{Correlation}
\label{corr}

Our primary observable obtained from the ASC correlator is the interferometric visibility $\tilde V$ in the domain of frequency, $\nu$ and time, $t$. This is the product of the Fourier transforms of electric fields at two antennas $A$ and $B$, averaged over a time interval large in comparison to the inverse of the Nyquist sampling rate, with
\begin{equation}
\tilde V_\mathrm{AB} (\nu ,t) =\left\langle\tilde E_A(\nu ,t) \tilde E^*_B (\nu , t)\right\rangle .
\label{eq:tildeV}
\end{equation}
$\tilde V_\mathrm{AB} (\nu ,t)$ is also known as the cross-power spectrum or cross spectrum or dynamic cross spectrum.  In Paper I we show how the visibility $V$ can be represented in four domains given by the four combinations of $\nu$, $t$, $\tau$, and $f$, where the latter two parameters are delay and fringe rate \citep[see also][]{brisken2010}. In this paper we are concerned with the visibility in the domain of delay and fringe rate, $V_\mathrm{AB} (\tau, f)$, from which we derive the size of the scatter-broadened image of the pulsar. This is the two-dimensional inverse Fourier transform of $\tilde V_\mathrm{AB} (\nu ,t)$ with
\begin{equation}
V_\mathrm{AB} (\tau, f) = {\mathfrak F}^{-1}_{\nu\rightarrow \tau}\left[  {\mathfrak F}_{t\rightarrow f}\left[ \tilde V_\mathrm{AB} (\nu, t) \right] \right].
\label{eq:V}
\end{equation}

To provide maximum sensitivity for the correlation, the data were correlated only during the ON-pulse time interval of the average pulse profile. We determined the phase of this interval, or window, within the pulsar period from the autocorrelation spectra (auto-spectra), $\tilde V_\mathrm{AA} (\nu, t)$, computed for each of the three ground stations, including GBT, and SRT.
This is the product of electric fields in the domain of frequency $\nu$ at antenna $A$ where $A$ stands for any of the four antennas:
\begin{equation}
\tilde V_\mathrm{AA} (\nu ,t) =\left\langle\tilde E_A(\nu ,t) \tilde E^*_A (\nu , t)\right\rangle .
\label{eq:VAA}
\end{equation}
Also,
\begin{equation}
V_\mathrm{AA} (\tau, t) = {\mathfrak F}^{-1}_{\nu\rightarrow \tau}\left[ \tilde V_\mathrm{AA} (\nu, t) \right].
\end{equation}

The correlator output was sampled synchronously with the pulsar
period of 0.714~s (single pulse mode). We used the ephemerides
computed with the program TEMPO for the Earth center. Integration over time yielded the average pulse profile
and the phase of the ON-pulse window.

\begin{figure}
\centering
\includegraphics[angle=270,trim=0cm 0cm 0cm 0cm,width=0.48\textwidth]{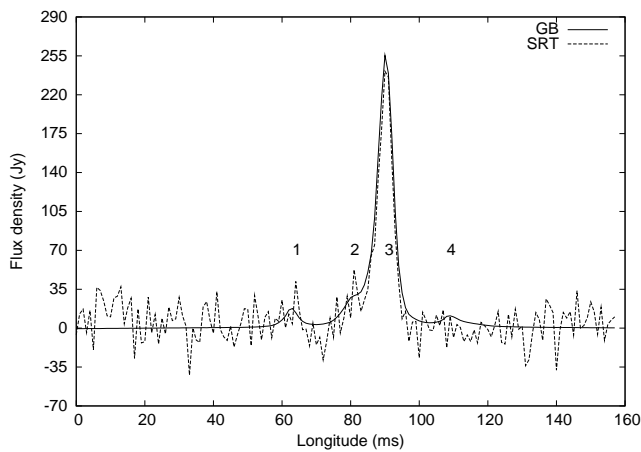}
\caption{The average pulse profile of PSR B0329+54 observed with the GBT and the SRT. The four components of the pulsar are indicated by numbers. The main component is indicated by number 3. The ON-pulse window was centered on this component. The pulsar period is equal to 714~ms.
}
 \label{fig:Prof}
\end{figure}

The ON-pulse window was centered on the main component of the
average profile, with a width of 8~ms. We also chose an OFF-pulse
window, offset from the main pulse by half a period and with the
same width as the ON-pulse window. With the phase and duration of
the windows determined, all VLBI data were correlated with gating
and dedispersion applied using 2048 channels. The correlator
computed the cross-spectra, $\tilde V_\mathrm{AB} (\nu ,t)$ which were then
written in standard FITS format for further analysis.

\subsection{Calibration}
\label{cal}

\subsubsection{Determining the SEFD values for the telescopes}

The amplitudes of the average profiles allowed us to accurately determine the sensitivities of all radio telescopes including SRT. 
The GBT on-source system temperature was monitored during the observations. It was measured to be about 140~K. With the gain value of 2.0~K/Jy (Green Bank Observatory Proposer's Guide for the Green Bank Telescope\footnote{\url{https://science.nrao.edu/facilities/gbt/proposing/GBTpg.pdf}}), we derive on-source $SEFD^\mathrm{GBT}_\mathrm{psr}=70$~Jy.
The peak flux density of the pulsar in a selected scan was therefore estimated as $\Delta S$=233~Jy. For the SRT and the same scan, $\Delta S/SEFD^\mathrm{SRT}_\mathrm{psr}$ was found to be 0.008.
After correction for the one-bit sampling factor of 1.57, this ratio becomes 0.0125,
yielding $SEFD^\mathrm{SRT}_\mathrm{psr}=18,650$~Jy. This is in a good agreement with the $SEFD^\mathrm{SRT}$ =19,000 Jy measured for SRT in the total
power mode in space by \cite{kovalev2014}.
Similar estimates give on-source $SEFD^\mathrm{WB}_\mathrm{psr}=255$~Jy and  $SEFD^\mathrm{KL}_\mathrm{psr}=650$~Jy for WB and KL, respectively.
In fact, we do not use these values in our calculations, since we analyse normalized visibility amplitudes as described in the next section. The normalization factor $R$ consists of parameters measured directly from correlator raw cross-spectra and auto-spectra for ON-pulse and OFF-pulse windows.
We estimated $SEFD$ for participating telescopes and the illustration in Figure~\ref{fig:Prof} in order to show that even the 10-m space radio telescope provides sufficient sensitivity to obtain normalized visibility amplitudes with a high  accuracy.

\subsubsection{Calibration of the visibility data}

For the subsequent analysis we used the CFITSIO package
\citep{pence1999}. We modified the visibility data in three steps.
First, we corrected for strong narrow-band interference by replacing
the affected data in a scan with properly scaled random numbers.
Second, we applied corrections for the SRT, GBT, KL and WB receiver
bandpasses using OFF-pulse auto-spectra as a template.  Third, we
used OFF-pulse auto-spectra to take into account strong intensity
fluctuations of pulsar radio emission caused by both intrinsic
variability of the pulsar emission and scintillation of that emission in the ISM, as described below.
These fluctuations affect the amplitude of the visibility function. In cases where corrections were applied to single pulses, we normalized the cross-spectra by using the normalization factor $R_\mathrm{cor}$:
\begin{equation}
R_\mathrm{cor}^{-1}=\sqrt{(\sigma_\mathrm{1on}^2-\sigma_\mathrm{1off}^2)(\sigma_\mathrm{2on}^2-\sigma_\mathrm{2off}^2)}
\label{eq:R1}
\end{equation}

Here, $\sigma_\mathrm{1on}^2$ and $\sigma_\mathrm{1off}^2$ are the
variances for the pulses observed at the ground radio telescopes used for cross-correlation,
computed in ON-pulse and OFF-pulse windows as
the sum of all harmonics in the power auto-spectra. Similarly,
$\sigma_\mathrm{2on}^2$ and $\sigma_\mathrm{2off}^2$ are the corresponding variances for the SRT
spectra computed in the same way.

The relatively low sensitivity of SRT renders the measurement of
$\sigma_\mathrm{2on}^2-\sigma_\mathrm{2off}^2$ rather uncertain. It can even give
negative values. We therefore replaced  this ratio with the ON-pulse
intensity for a given ground radio telescope (GRT), but reduced to
an expected value for the SRT using the coefficient $\eta =
SEFD_\mathrm{GRT}/SEFD_\mathrm{SRT}$.  This coefficient was taken to be equal to
0.0024, 0.0063, and 0.016 for GBT, WB and KL respectively,
which, with the sampling factor taken into account, correspond to
the ratios of the SEFD values given in section~\ref{obs}.  Then,
equation~\ref{eq:R1} can be written as
\begin{equation}
R_\mathrm{cor}^{-1}=\frac{\sigma_\mathrm{2off}}{\sigma_\mathrm{1off}}(\sigma_\mathrm{1on}^2-\sigma_\mathrm{1off}^2)\sqrt{\eta}\,.
\label{eq:R2}
\end{equation}
Instead of corrections to individual cross-spectra $\tilde V_\mathrm{AB} (\nu,t)$
we applied the corrections to visibility amplitudes $V^\mathrm{amp}_\mathrm{AB}$
in the following way. For 1170-s scan we calculated 16 raw complex
visibility functions $V_\mathrm{AB} (\tau, f)$ for every 100 pulses (array
dimension 2048$\times$100). Then we took the visibility amplitude as
$V^\mathrm{amp}_\mathrm{AB}=\sqrt{R^2_\mathrm{AB}+I^2_\mathrm{AB}}$ at a point of maximum (close
to $\tau$=0 and $f=0$) as it is used in any VLBI study (R and I being
real and imaginary components of complex function). We applied the
same routine to the auto-spectra ON-pulse and OFF-pulse longitudes:
computing $V^\mathrm{amp}_\mathrm{AA}(\mathrm{ON})$, $V^\mathrm{amp}_\mathrm{AA}(\mathrm{OFF})$,
$V^\mathrm{amp}_\mathrm{BB}(\mathrm{ON})$, and $V^\mathrm{amp}_\mathrm{BB}(\mathrm{OFF})$. These quantities are equivalent to the corresponding variances $\sigma_\mathrm{1on}^2$,
$\sigma_\mathrm{1off}^2$, $\sigma_\mathrm{2on}^2$ and $\sigma_\mathrm{2off}^2$, averaged over integration time $\Delta t_\mathrm{int}=71.4$~s. Then we corrected raw
values of $V^\mathrm{amp}_\mathrm{AB}$ using equation~\ref{eq:R1} for WB-KL
baseline combination, and using equation~\ref{eq:R2} for WB-SRT and
KL-SRT baseline combinations with the coefficient $\eta$ mentioned
above. We have averaged 16 corrected values of visibility amplitude
$V^\mathrm{amp}_\mathrm{AB}$ for every 1170-s observing scan, and these values are
displayed in Figure~\ref{fig:VisDist} as a function of projected
baseline length.

\section{Visibility as a function of projected baseline length}
\label{IntVis}

Important parameters for the subsequent analysis are the scintillation time $t_\mathrm{diff}$ and the decorrelation bandwidth $\Delta\nu_\mathrm{diff}$. We have estimated these parameters by computing the two-dimensional cross-correlation function (CCF) between the ON-pulse dynamic auto-spectra obtained over 1170-s scans at KL and WB. More specifically, we used sections of the CCF along time and frequency delay axes, fitted with Gaussian and exponential functions respectively. The parameters, $t_\mathrm{diff}$ and $\Delta\nu_\mathrm{diff}$ are listed in Table~\ref{tab:scatt_params}.

\begin{table}
\centering

	\caption{Pulsar scattering parameters}
	\label{tab:scatt_params}
	\begin{tabular}{c c c c} 
	\hline
	\hline
	Epoch  	& $t_{\rm diff}$	& $\Delta\nu_{\rm diff}$	&	$\tau_\mathrm{scat}$	\\
			& (s)                	& (kHz)		                &	($\mu$s)			\\
	(1) 	 	& (2)               	& (3)	                         	&(4)				\\
	\hline
	2013  November 22      & $110\pm2$               & $7\pm 2$                      & $11.6\pm0.8$    \\
	2014  January  1, 2    & $102\pm2$               & $7\pm2$                       & $12.5\pm0.7$       \\
	\hline
	\end{tabular}

\raggedright
Note: Columns are as follows: 
(1) date of observations;
(2) scintillation time from single-dish autocorrelation spectra as the half width at 1/e of maximum;
(3) scintillation bandwidth from single-dish autocorrelation spectra as the half-width at half maximum (HWHM);
(4) HWHM of an exponential function fit to the visibility amplitude distribution along the delay axis.
\end{table}

For our 2013~November and 2014~January observing
sessions each visibility function $V_\mathrm{AB} (\tau, f)$ was computed
over 100 pulsar periods ($\Delta t_\mathrm{int} = 71.4$~s), short enough so
that no coherence losses were expected and no correction for it
needed to be applied.
This time is also smaller than the scintillation time, $t_\mathrm{diff}$, and therefore small enough to be in the so-called snapshot regime
\citep{goodman1989} where no damping of
visibility amplitudes is expected.

\begin{figure}
\centering
\includegraphics[angle=0,trim=0cm 1cm 0cm 0cm,width=0.48\textwidth]{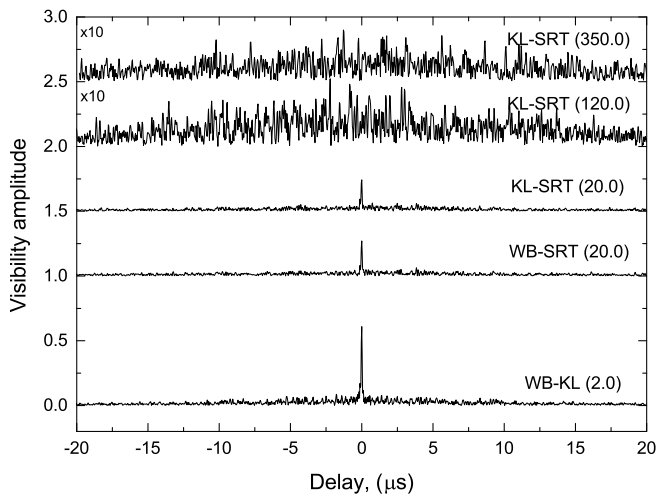}
\caption{
Examples of the cross section, $|V(\tau, f_\mathrm{max})|$, of the distribution of the visibility magnitude $|V(\tau, f)|$ at the fringe rate, $f_\mathrm{max}$ with maximum magnitude. This fringe rate is near zero. The upper curves correspond to the longest baseline projections and the lower curves to correspondingly shorter
baselines, with the telescopes indicated and the approximate
baseline projections given in M$\lambda$ in parentheses. The four
upper curves are offset by constants 1.0, 1.5, 2.0, and 2.5 for ease of viewing and the two top curves are, in addition, magnified by a factor 10.
\label{fig:delay_fine_structure}}
\end{figure}

In Paper~I we showed an example of the distribution of visibility as
a function of delay and fringe rate, $|V(\tau, f)|$, for a
ground-space baseline. In Figure~\ref{fig:delay_fine_structure} we
present cross sections, $|V(\tau, f_\mathrm{max})|$, of the distributions
for four different baselines, each for a fixed fringe rate near zero
where the maximum of the distribution occurred. The curves are
plotted according to baseline length, with the shortest baseline of
2 M$\lambda$ at the bottom of the figure and with the longest
baseline of 120 M$\lambda$ at the top.

It is apparent that the highest visibility amplitude occurs at the
shortest baseline with one spike near zero delay. This spike is due
to the scatter-broadened image of the pulsar.  The two curves at
intermediate baselines of 20 M$\lambda$ length still show the spike
at zero delay but at diminished strength. The pulsar appears to be
partly resolved on these baselines. The top curve at 300 M$\lambda$
baseline length does not show a pronounced spike anymore. The
scatter-broadened image of the pulsar is completely resolved. Only a
broad distribution of visibility magnitudes remains. This
distribution gives unique information about the scattering
properties of the interstellar medium along the line of sight to the
pulsar.

\begin{figure}
\centering
\includegraphics[angle=0,trim=0cm 1cm 0cm 0cm,width=0.49\textwidth]{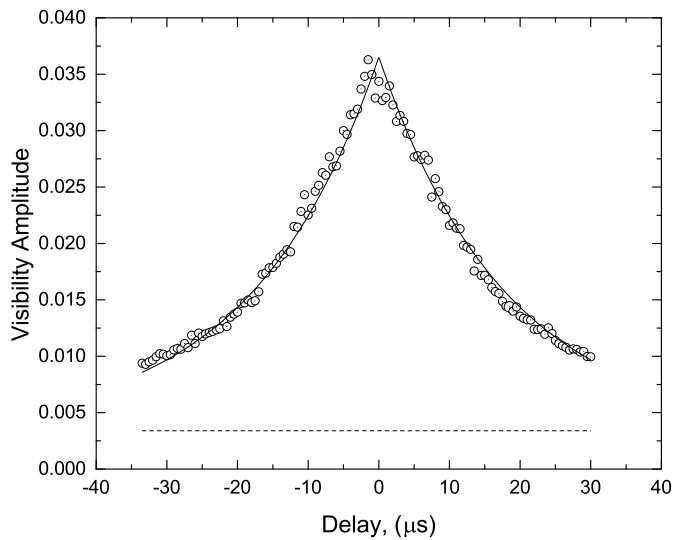}
\caption{An example for the KL-SRT baseline of a cross section,
$|V(\tau, f_\mathrm{max})|$,  at $f_\mathrm{max}$ where the magnitudes peaked,
namely close to zero mHz. The data were obtained on 2014 January 2. 
Individual visibility functions $V_\mathrm{AB}(\tau,f)$ were obtained
over 71.4 s, and then 16 functions were averaged. Additional
averaging has been done along the delay axis  by 16 points ($0.5\mu$s)
to smooth deep modulation. The solid line gives the exponential fit
and the dashed line the background level.
\label{fig:averdelay}
}
\end{figure}

To determine the visibility magnitudes as a function of projected
baseline length, we took the maximum value above the background
level in every visibility function, $V_\mathrm{AB}(\tau,f)$ (Eq.~\ref{eq:V})
both for the short baselines where a spike at about zero delay and
fringe rate appeared, and for the long baselines where the maximum
spike can appear at arbitrary delay and fringe rate.
To determine a scattering time we used the visibility functions $V_\mathrm{AB}(\tau,f)$ obtained for the longest baseline projections where isolated spikes did not
dominate. Figure~\ref{fig:averdelay} gives  an example of such a visibility 
function for a fixed fringe rate, $f_\mathrm{max}$ close to zero Hz and averaged for ease of displaying the functional dependence. 
The visibility function was obtained for the
KL-SRT baseline on 2014 January 2 for the single observing scan. A total of 16
visibility functions, each obtained over 100 pulses (71.4 s), were
averaged. Additional averaging was done along the delay axis by
16 points (0.5 $\mu$s) for further smoothing. Then
we fit an exponential function with a background level to the
distribution and determined the scattering time $\tau_\mathrm{scat}$ from the
fit. It is also listed for the two epochs in Table~\ref{tab:scatt_params}.
Figure~\ref{fig:averdelay} shows the averaged cross section of the visibility function and the exponential fit.

For the determination of the visibility amplitude as a function of
baseline length we took the maximum value in each visibility
function as was described in section~\ref{cal}.

\begin{figure}
\centering
\includegraphics[angle=0,trim=0cm 0cm 0cm 0cm,width=0.47\textwidth]{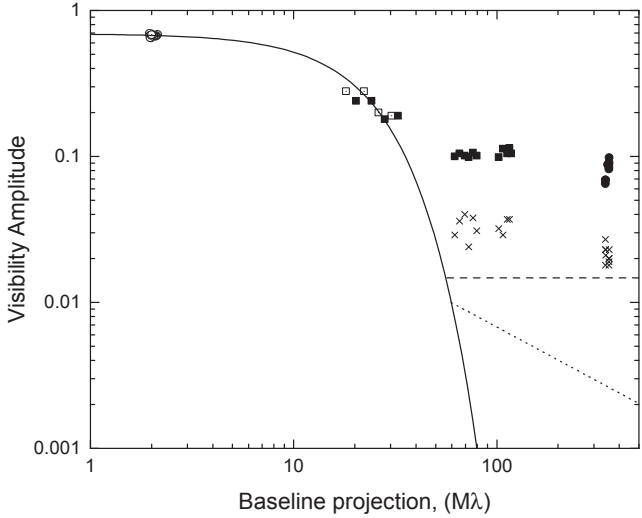}
\caption{
Visibility amplitude versus projected baseline length in millions of wavelengths. The visibility amplitudes were obtained in two different ways. First, they were computed as the maxima, $|V_\mathrm{AB}(\mathrm{max})|$, in the $|V(\tau, f)|$ distributions for each baseline AB as follows. Open circles: WB-KL baseline on 2014 January 1; open squares: WB-SRT baseline on 2014 January 1; filled squares: KL-SRT baseline on 2014 January 1, 2; filled circles: KL-SRT baseline on 2013 November 22. Note, the visibility amplitudes on baselines longer than 60 M$\lambda$ do not in general correspond to the amplitudes at near zero delay and fringe rate. Second, for the long baselines visibility amplitudes were also computed as the maxima of fit exponentials to the averages of many cross section $|V(\tau, f_\mathrm{max})|$ visibilities (see text in the section~\ref{IntVis}): crosses: KL-SRT baseline on 2014 January 1,2 (60 to 120 M$\lambda$) and KL-SRT baseline on 2013 November 22 (about 350~M$\lambda$ corresponding to 330,000 km). The solid line represents the fit by Eq.~\ref{eq:theta_B}, corresponding to a scattering disk size of 4.8~mas. Dashed line denotes expected level of diffractive substructure.
The dashed line for diffractive substructure corresponds to the estimate in section~\ref{sec:comparison} (a fractional amplitude of 0.021), while the refractive model is given in Eq.~\ref{eq:mike}, and is described by \citet{johnson2015}.
}
\label{fig:VisDist}
\end{figure}

These values were then averaged for every 16 $\times$100-pulsar periods, that is for 1170 s, and they are
displayed in Figure~\ref{fig:VisDist} as $|V_\mathrm{AB}(\mathrm{max})|$ with open circles and open and filled squares for different baselines. 
However, while for projected baselines up to about 40~M$\lambda$ $|V_\mathrm{AB}(\mathrm{max})|$ occurred at or close to the origin of zero delay and fringe rate, for baselines longer than about 40~M$\lambda$ $|V_\mathrm{AB}(\mathrm{max})|$ occurred in general considerably away from the origin (see, Figure~\ref{fig:delay_fine_structure}).
To determine $|V_\mathrm{AB}|$ at the origin for the long baselines to be comparable to $|V_\mathrm{AB}(\mathrm{max})|$ at the shorter baselines, fluctuations had to be reduced. We had first computed the average of the cross section $|V(\tau, f_\mathrm{max})|$ visibilities for 1170~s as displayed in Figure~\ref{fig:averdelay} for just one scan, but subsequently did it for all such long-baseline scans. 
We then fit an exponential function to the averages and determined the peak visibility 
magnitudes which were always close to the origin.
These averages are equal to the rms values over time of the scans since the visibilities are $\chi^2$~distributed with two degrees of freedom. 
In this sense, they are important for comparisons with theoretical predictions.

There are four groups of points in Figure~\ref{fig:VisDist}: visibility amplitudes on
short, $\sim$2~M$\lambda$ ground-ground baselines, on intermediate, 15 to 40  M$\lambda$ and long, 60 to 120 M$\lambda$,
ground-space baselines, and on the longest, about 350~M$\lambda$, baselines. We estimate the $1\sigma$ uncertainties of the visibility amplitudes to be about as large as the symbols. Larger variations may be caused by the varying baseline and/or by scintillation.

The expected visibility amplitude as a function of baseline length due
to angular broadening is given by \citet{gwinn1988}, apart from a scaling factor, as
\begin{equation}
V_\mathrm{AB}=\exp\left\{-\frac{1}{2}\left[\frac{\pi}{(2\ln{2})^{1/2}}\frac{\theta_{H}b}{\lambda}\right]^{\alpha-2}\right\}
\label{eq:theta_B}
\end{equation}
with, e.g., $\alpha=4$, for a Gaussian and where $\theta_{H}$ is a full width at half maximum (FWHM) of a Gaussian image of scattering disk. The parameter, $\alpha$, is
the spectral index of electron density fluctuations in the ISM, and $b$ is the projected baseline of the interferometer. It is clear that this function
cannot provide a good fit to the visibility amplitudes for all
baselines. Only those on the short and intermediate baselines
provided a good fit with the fitting function shown as a solid line.
The power index $\alpha$ for PSR~B0329+54 was measured by
\citet{shishov2003} in their multi-frequency study; they found
$\alpha=3.50\pm 0.05$. We used this value for our fit.

We obtained $\theta_{H}=2.4\pm0.4\times 10^{-8}$~rad or $4.8\pm0.8$~mas from the best fit of Eq.~\ref{eq:theta_B} to the data for shorter baseline projections below 50~M$\lambda$.
The visibility amplitudes at zero baseline are, as expected, smaller than unity since
$\Delta\nu_\mathrm{int}$ is much larger than $\Delta\nu_\mathrm{diff}$ and we are
therefore integrating in frequency over many scintels, i.e. regions of increased power in the cross-spectrum.
Our fit function (Eq.~\ref{eq:theta_B}) falls toward zero very rapidly for baselines longer than 60~M$\lambda$. Indeed, the function is many orders of magnitude smaller at the longest baselines than the visibility amplitudes we measured. The latter are approximately constant, varying only between about 0.11 and 0.07 for baselines from 60 to 350~M$\lambda$. Clearly the scatter-broadened image of PSR~0329+54 is resolved and substantial substructure exists in it.

\section{The scattering material in the ISM along the line of sight to the pulsar} \label{scat}

At a parallax distance of $D=1.03^{+0.13}_{-0.12}$~kpc \citep{brisken2002}, PSR~0329+54 is located at the outer edge of the Orion spiral arm, while the Sun is
close to the inner edge of the arm.
A comparison of angular and temporal broadening provides information on the distribution of the scattering material in this arm along the line of sight to the pulsar. Expressions for the angular-broadened diameter $\theta_H$ (FWHM) and the broadening time $\tau_\mathrm{scat}$ have been derived by \citet{blandford1985}. These expressions contain the function $\psi(z)$ which defines the mean scattering angle per unit length, $z$, measured from the source to the observer. 
For our study we take $\tau_\mathrm{scat}$ as the mean of the two values listed in Table~\ref{tab:scatt_params}.
If we assume the extreme case that the scattering material is uniformly distributed along the line of sight to the pulsar, then $\psi(z)$ is a constant \citep{britton1998} and
$\theta_H^2=16\ln{2}(c\tau_\mathrm{scat}/D)$ with $D$ being the distance to the pulsar. This relation gives for our values of the broadening time, $\tau_\mathrm{scat}$ of
$12.1\pm0.5~\mu$s  value of $\theta_H=(3.6\pm0.1)\times 10^{-8}$~rad which is significantly larger than the measured value of $\theta_H=(2.4\pm0.4)\times 10^{-8}$ rad. Therefore, for this relatively simple case, a model of uniformly distributed scattering material along the line of sight to the pulsar is inconsistent with our result at a level of $3\sigma$.

If we assume another extreme case that the scattering material is only distributed in a thin screen located at a distance $d$ from Earth, then $\psi$ becomes a function
of $z$, $D$, and $d$ and $\theta_H=[8 \ln{2} c\tau_\mathrm{scat} (D-d)/Dd]^{1/2}$
\citep{britton1998}. This relation gives us for our measured values of $\theta_H$ and
$\tau_\mathrm{scat}$ from the January session, $d=(0.5\pm0.1)D$ which means that the
effective screen is located approximately in the middle of the distance
to the pulsar.

However, \citet{shishov2003} suggested that interstellar plasma along the line of sight to PSR~B0329+54 consists of two types of inhomogeneities: turbulent irregularities, which produce the diffractive scattering, and large scale irregularities, which produce the frequency dependent angular refraction. In this more realistic case a different relation between $\theta_H$ and the distances, $D$ and $d$, needs to be considered. This relation includes the diffractive scintillation time, $t_\mathrm{diff}$, the associated diffractive length-scale \citep[see for definition][]{narayan1989} in the plane of the observer normal to the line of sight to the pulsar, $r_\mathrm{diff}$, and the velocity of the scintillation pattern with respect to the observer, $v_\mathrm{diff}$, with
\begin{equation}
r_\mathrm{diff}=v_\mathrm{diff}\,t_\mathrm{diff}
\label{eq:rvt}
\end{equation}
The velocity, $v_\mathrm{diff}$, is a geometrical function of the velocities of the observer, $v_\mathrm{obs}$,
the scattering screen, $v_\mathrm{scr}$, and the pulsar, $v_\mathrm{psr}$, as well as the distances, $D$ and $d_\mathrm{s}$,
and is given by
\begin{equation}
v_\mathrm{diff}=v_\mathrm{obs}-\frac{D}{d_\mathrm{s}}v_\mathrm{scr}+\frac{D-d_\mathrm{s}}{d_\mathrm{s}}v_\mathrm{psr}
\label{eq:vdiff}
\end{equation}
Here $d_\mathrm{s}$ is a distance from the pulsar to the screen.
We estimated $r_\mathrm{diff}$ by comparing the expression for the visibility amplitude versus baseline, given by \citep{goodman1989,prokhorov1975}
\begin{equation}
V_\mathrm{AB}=\exp[-\frac{1}{2}(\frac{b}{r_\mathrm{diff}})^{\alpha-2}]
\label{eq:Vr_diff}
\end{equation}
with equation~\ref{eq:theta_B}. We obtained
\begin{equation}
r_\mathrm{diff}=(\frac{\lambda}{\theta_H\pi})\sqrt{2\ln{2}},
\label{eq:r_diff}
\end{equation}
and $r_\mathrm{diff}=(1.7\pm0.3)\times10^4$ km, with the error derived from
the error of $\theta_H$. The mean value of $v_{diff}$ is 170~km/s.
The pulsar velocity was taken from \citep{brisken2002} as $v_\mathrm{psr}$=90~km/s We assume $v_\mathrm{scr}$=0, and $v_\mathrm{obs}$=34.6~km/s for January 1 2014. With these values we get $\frac{d_\mathrm{s}}{D}=0.37\pm0.06$ or  $\frac{d}{D}=0.63\pm0.06$. 
This estimate is somewhat larger than the single screen model value of $\frac{d}{D}=0.50\pm0.1$ but still consistent with it within it the combined error. Averaging the two results we get $\frac{d}{D}=0.60\pm0.1$.

\section{Substructure in the scatter-broadened image of PSR 0329+54}
\label{subst}

\subsection{Visibility amplitudes as a function of projected baseline length}

The diffraction scatter-broadened image of PSR~0329+54 with angular size $\theta_H$ is reflected in the rapidly decaying visibility amplitudes
up to $b\sim40$~M$\lambda$. For longer baselines the image should be completely resolved, and on the basis of the fitting function (Eq.~\ref{eq:theta_B})
no visibilities expected to be detected. However, visibilities are detected, and, while there is a scatter in the measured values of the visibility amplitudes, the values are approximately constant up to the longest baselines of $\sim$350~M$\lambda$. Further, as shown in Figure~\ref{fig:delay_fine_structure}, the cross sections of the secondary spectra are
qualitatively different for different baselines. For shorter baselines, the cross sections consist of a broad component and a central narrow spike whereas
for long baselines the central narrow spike disappeared and only a broad distribution of low amplitude spikes along delay remains. How can that be understood?

It is important to consider general aspects of the ground-space interferometer and the scattering medium.
For radio waves through the ISM, scattering is considered to be strong, meaning that the lengths of the many paths the scattered radio waves take from the pulsar to the observer differ by many wavelengths. \citet{goodman1989} and \citet{narayan1989} distanced three cases for strong scattering of which two are of relevance for our space VLBI data, the snapshot and the average regime. In the snapshot regime, the relative phases of the radio waves travelling along different paths remain mostly unchanged during the observations. The interferometer observes a pattern of
speckles whose parameters are a function of the size of the source and of small and large-scale electron density variations in the ISM.
The average regime is characterized by some averaging over several different paths smoothing out to zero visibility due to small-scale density variations and only leaving the visibility from the large-scale variations.

For our observations, the size of the source is very small. Pulsar
radio emission is generated in a very compact region inside the
magnetosphere of a neutron star. Even if the size of the
emission region is as large as the light cylinder, the angular size
of that region for PSR 0329+54 would be only $2.3\times 10^{-12}$
rad or 0.5~$\mu$as, smaller than 0.001 times the angular resolution
of VLBI with RadioAstron at 324 MHz on any baseline. Therefore, pulsars are
essentially ``point-like'' radio sources to interferometers even 
on ground-space baselines, and all signatures of the
visibility data are therefore due to the small and large-scale
density variations in the ISM. Further, the averaging of our cross
spectra over 100 pulse periods or 71.4 s, is less than $t_\mathrm{diff}$
and therefore small enough to be in the so-called snapshot regime
where no damping of visibility amplitudes is expected. However, the
receiver bandwidth, $\Delta\nu_\mathrm{int}$ of 16~MHz is much larger than
the decorrelation bandwidth $\Delta\nu_\mathrm{diff} \sim 7$~kHz,
resulting in a decrease of the visibility amplitudes.

In our observations the $b/r_\mathrm{diff}$ ratio varies from 0.1 to
about 15. For a small ratio (up to about 1) the projected baseline was small and the corresponding
beam size of the interferometer averages
over many paths the scattered radio waves took from the pulsar to Earth. The ground telescope and
the SRT were clearly still in a diffraction spot. This is the average regime where the visibilities
from any possible small-scale density variations of the ISM were averaged to zero and only those
from large-scale density variations survived. On these baselines our interferometer only measured
the scatter-broadened image of the pulsar. For larger ratios of $b/r_\mathrm{diff}$ (up to 15) the two
telescopes were not
in the same diffraction spot anymore. The scatter-broadened image was completely resolved resulting
in essentially zero visibility. However, any small scale variations if they existed would be less
prone to averaging and could become detectable. Indeed, at baseline projections greater than
$r_\mathrm{diff}$ significant visibilities were detected and their amplitudes were scattered around a mean of 10 to 15\% 
of the maximum at short baselines
that remained approximately constant up to the
largest ground-space baselines of 330,000~km. Clearly, substructure in the scatter-broadened image
of PSR 0329+54 was detected, most likely due to small-scale electron density fluctuations in the
ISM.

\subsection{Comparison with Theory and Other Observations}
\label{sec:comparison}

How does the level of the rms visibility amplitudes around the mean of 3 to 5\% of the maximum at short baselines 
compare with predictions from theoretical considerations and how does it compare with other observations?
Substructure in scattering disks falls into two separate regimes: diffractive and refractive. These two regimes correspond to different scales of fluctuations in electron density; diffractive effects arise from fluctuations on scales of $r_\mathrm{diff} \sim 10^4$~km, while refractive effects arise from fluctuations on scales of $D \theta_H \sim 10^9$~km. These two dominant scattering scales can arise from a single power-law spatial spectrum for density fluctuations, such as the Kolmogorov spectrum \citep{rcb_1984}.

Diffractive substructure produces random visibility fluctuations of order unity that decorrelate rapidly in frequency and time (quantified by the scintillation bandwidth and timescale) \citep{gwinn2001,johnson_gwinn_2013}. Refractive substructure produces visibility fluctuations with smaller amplitudes, but these fluctuations have a decorrelation bandwidth of order unity and persist for weeks to months \citep{narayan1989,goodman1989,johnson2015}. Because of these properties, diffractive scintillation disperses source power in the delay-rate domain while refractive scintillation contributes power to a single peak (just as the unscattered source visibility does).  

For an effectively pointlike source such as a pulsar, the expected level of diffractive substructure on long baselines depends on the number of averaged scintillation elements. Our 71.4~s averaging timescale is shorter than the scintillation timescale in both epochs but our averaged bandwidth (16~MHz) is much wider than the 7~kHz scintillation bandwidth (see Table~\ref{tab:scatt_params}). We therefore expect an rms visibility amplitude on long baselines from diffractive scintillation of approximately $\sigma_\mathrm{diff} \approx \sqrt{(16~{\rm MHz})/(7~{\rm kHz})} = 0.021$, where this value corresponds to the fractional amplitude relative to the zero-baseline visibility.  

The expected level of refractive substructure on long baselines depends on the scattering properties and also on the baseline length $b$ but does not depend on the averaging timescale or bandwidth. Taking $\alpha=3.5$ \citep{shishov2003} and also using our derived values of the scattering, $d$ and $\theta_H$, we then find a fractional rms visibility from refractive substructure of \citep{johnson2015}:
\begin{align}
\sigma_\mathrm{ref}(b) = \sqrt{\langle | \Delta V_\mathrm{AB} |^2 \rangle}  &= 0.0091 \left( \frac{b}{10^5\ {\rm km}} \right)^{-0.75}\!.
\label{eq:mike}
\end{align}
For the range of baselines in Figure~\ref{fig:VisDist} that resolve the scattering disk, $b = 65$ to $350\ {\rm M}\lambda = 60{,}000$ to 280,000~km, we expect $0.013 \le \sigma_\mathrm{ref}(b) \le 0.005$.

Comparing these values, we see that our detected visibilities on long baselines are roughly consistent with the expected properties and strength of diffractive substructure. Moreover, our non-detections of strong, persistent peaks in delay-rate space on long baselines are also consistent with the expected strength of refractive substructure. Our observations of PSR~B0329+54 are the first to detect the signatures of diffractive substructure on baselines that entirely resolve the ensemble-average scattered image. And while refractive substructure has been detected in the galactic center radio source Sgr~A$^*$~\citep{gwinn2014} and in the quasar 3C\,273 \citep{kovalev2016,johnson2016}, our results are the first detections of either class of substructure that are not sensitive to properties of the intrinsic source. 
If the visibilities could be coherently integrated indefinitely, then as the integration time $T_\mathrm{int}$ increases the diffractive power in the delay-rate spectrum (e.g., Figure~\ref{fig:delay_fine_structure}) should decrease as $1/\sqrt{T_\mathrm{int}}$. With sufficiently long integration (but shorter than the refractive timescale), a persistent single spike would emerge, concentrated at a single location, containing the refractive power.

\section{Summary and conclusions}\label{Concl}

Here we summarize our observations and results and give our conclusions.
\begin {enumerate}
\item We made VLBI observations of PSR B0329+54 with RadioAstron at 324 MHz on projected baselines up to 330,000~km or 350~M$\lambda$. Our goal was to investigate scattering properties of the ISM which affect radio observations of all celestial sources. 
While the results of such observations are in general influenced by
the convolution of source structure with the scattering processes,
pulsars are virtually point-like sources and signatures in the
observational results can be directly related to the ISM scattering
properties.

\item Visibility function at short ground-ground baselines manifests a single bright spike in delay-rate
space that vanishes on long space-ground baselines. Thus, the scattering disk of PSR B0329+54 
was completely resolved on ground-space baselines of 15,000 to 30,000 km. 
The FWHM of the angular diameter is $4.8\pm0.8$~mas at 324~Hz.

\item The diffractive length scale or size of the diffraction spot near Earth is
$17,000\pm3,000$~km.

\item With the assumption of turbulent and large-scale irregularities in the plasma, the effective scattering screen is located at $d/D=0.6\pm0.1$ or somewhat more than half of the distance from Earth to the pulsar.

\item At longer projected baselines, up to 330,000 km, significant visibility amplitudes were detected, although none were expected from the scattering disk. They are scattered around a mean which stays approximately constant up to the longest baselines. This result indicates that substructure was discovered in the scatter-broadened image of PSR~B0329+54.
\end{enumerate}

\section*{Acknowledgements}

We thank the anonymous referee for thorough reading of the manuscript and valuable comments which helped to improve the paper.
The RadioAstron project is led by the Astro Space Center of the Lebedev Physical Institute of the Russian Academy of Sciences and the Lavochkin Scientific and Production Association under a contract with the Russian Federal Space Agency, in collaboration with partner organizations in Russia and other countries. The National Radio Astronomy Observatory is a facility of the National Science Foundation operated under cooperative agreement by Associated Universities, Inc. This research was supported by Basic Research Program P-7 of the Presidium of the Russian Academy of Sciences.

{\it Facilities:} RadioAstron Space Radio Telescope (Spektr-R), GBT, WSRT, Kalyazin radio telescope.

\bibliographystyle{mnras}
\bibliography{0329_paper2}

\bsp	
\label{lastpage}
\end{document}